# Symmetry-Enforced Ferroelectric Switching of Two-Dimensional Altermagnetism

Jiangyu Zhao[1], Yibo Liu[1], Guoli Wu[1], Xinru Li[1], Ying Dai*[,1], Baibiao Huang[1], Yandong Ma*[,1]

[1]School of Physics, State Key Laboratory of Crystal Materials, Shandong University, Shandanan Street 27, Jinan 250100, China

*Corresponding author: daiy60@sdu.edu.cn (Y.D.); yandong.ma@sdu.edu.cn (Y.M.)

**ABSTRACT**

Altermagnetism features strong momentum-dependent spin splitting despite zero net magnetization, offering a transformative platform for next-generation spintronics. However, the nonvolatile and deterministic switching between its two equivalent spin-splitting states remains a fundamental bottleneck. Here, we propose a universal layer-engineering paradigm to achieve symmetry-enforced ferroelectric switching of two-dimensional altermagnetism. By sandwiching a conventional antiferromagnetic monolayer between two identical ferroelectric layers, the out-of-plane polarization cleanly breaks the spatial symmetry to induce robust altermagnetic splitting. Crucially, the global combined parity-time symmetry dictates that reversing the ferroelectric polarization exactly inverts the altermagnetic spin-splitting pattern. We rigorously validate this mechanism in the $In_2Se_3/MnPTe_3/In_2Se_3$ trilayer using first-principles calculations. As a direct consequence, the ferroelectrically driven spin-splitting reversal deterministically flips the anomalous Hall effect signal, providing an unambiguous transport fingerprint to electrically distinguish the two altermagnetic states. Unconstrained by the stringent symmetry requirements of intrinsic single-phase materials, our findings establish a versatile physical framework for electrically addressable altermagnetic spintronics.

## Introduction

Altermagnetism (AM) has recently emerged as a fundamentally distinct magnetic phase, characterized by strong momentum-dependent spin splitting despite a strictly zero net magnetization protected by crystal symmetries [1–7]. This unconventional spin ordering elegantly bridges the macroscopic advantages of antiferromagnets (e.g., absence of stray fields and ultrafast dynamics) with the momentum-space functionalities of ferromagnets, hosting exotic phenomena such as the anomalous Hall effect [1], the magneto-optical Kerr effect (MOKE) [8], and the current-induced spin polarization [9]. For practical applications in ultra-dense spintronic devices and high-speed magnetic memories [5,10], achieving nonvolatile and deterministic switching between the two equivalent and fully inverted AM spin-splitting states is a central prerequisite. Since the AM spin texture is rigidly locked to the crystal momentum, coupling it to a switchable spontaneous electric polarization—thereby achieving ferroelectric (FE) control of altermagnetism—has become a paramount challenge in the field.

Current efforts to conquer this challenge and establish practical electrical switchability face severe fundamental bottlenecks, broadly diverging into three directions. One intuitive approach is the search for intrinsic single-phase multiferroic altermagnets [11–14], where FE and AM orders coexist. However, this route demands exceptionally stringent symmetry prerequisites, severely restricting the pool of candidate materials and their practical tunability [11]. A second alternative circumvents these symmetry restrictions by artificially breaking the combined spatial-inversion and time-reversal ($PT$) symmetry of conventional antiferromagnets via external electric fields [1,15]. Yet, continuous electric-field gating suffers from intrinsic volatility. More recently, exploring sliding ferroelectricity in two-dimensional (2D) van der Waals heterostructures has emerged as a third, highly active platform [10,16–20]. Nevertheless, sliding mechanisms intrinsically yield very weak spontaneous polarizations and suffer from low energy efficiency compared to conventional out-of-plane ferroelectrics [21,22]. Consequently, developing a robust, highly efficient, and universally applicable paradigm to realize the nonvolatile electrical switching of AM states remains a critical and elusive frontier.

In this letter, we overcome these bottlenecks by proposing a universal layer-engineering paradigm that unlocks nonvolatile electrical control over 2D altermagnetism. Based on rigorous symmetry analysis, we demonstrate that sandwiching a conventional antiferromagnetic (AFM) monolayer between two identical intrinsic FE monolayers can lift the equivalence of the two magnetic sublattices within the AFM layer, thereby generating robust AM spin splitting. Crucially, because the upward and downward polarization states are strictly mapped onto each other via $PT$ operation, an electric-field reversal guarantees an exact, deterministic inversion of the entire altermagnetic band structure. Using the $In_2Se_3/MnPTe_3/In_2Se_3$ heterostructure as a prototypical platform, our first-principles calculations confirm this symmetry-enforced phase transition. We further demonstrate that this microscopic momentum-space reversal manifests macroscopically as a fully invertible anomalous Hall conductivity, serving as a highly distinct transport readout. By eliminating the reliance on rare single-phase multiferroics and overcoming the intrinsic weakness of sliding ferroelectrics, this architecture offers a highly practical and efficient pathway for next-generation altermagnetic spintronics.

## Results and Discussions

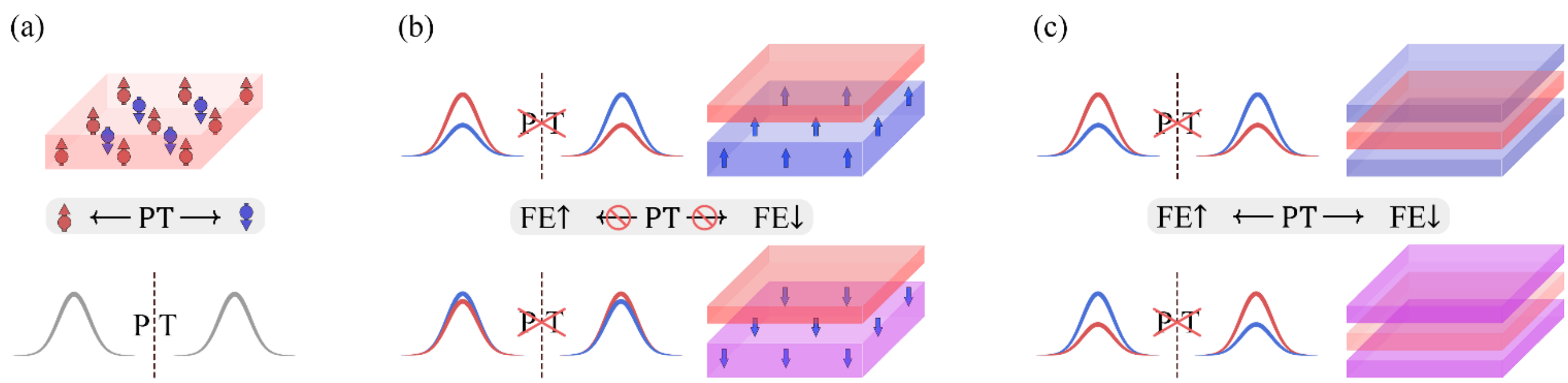

**Figure 1.** (a) Spin configuration and the corresponding spin-degenerate band structure (gray curves) of a generic collinear AFM monolayer, protected by $PT$ symmetry. (b) Spontaneous symmetry breaking in an asymmetric FE/AFM bilayer. The out-of-plane FE polarization (blue arrows) lifts the momentum-space spin degeneracy. However, the reversed FE state (bottom) lacks a symmetric mapping to the original state. (c) The proposed symmetric FE/AFM/FE trilayer architecture. The opposing FE polarizations (+z, blue; -z, purple) cleanly break spatial symmetry to generate AM splitting, while the global $PT$ symmetry strictly guarantees that reversing the FE polarization exactly inverts the spin-resolved band dispersions (red and blue curves).

We begin by establishing the underlying symmetry prerequisites for realizing AM in a generic two-dimensional antiferromagnet. In conventional collinear AFM systems, the magnetic sublattices are strictly connected by either $PT$ symmetry or combined translational and time-reversal ($\tau T$) symmetry. This symmetric arrangement, depicted in **Fig. 1(a)**, enforces identical effective potentials for opposite spin channels. Consequently, despite the microscopic staggering of magnetic moments, the local ligand-field anisotropy is fully compensated, preserving spin degeneracy across the entire Brillouin zone and rendering the system equivalent to a spin-isotropic model. To activate AM, this macroscopic $PT$ symmetry must be explicitly broken, allowing the highly anisotropic local crystal field of the individual sublattices to couple with the crystal momentum, thereby inducing a non-relativistic spin splitting. Simultaneously, a purely spatial-magnetic operation, such as a mirror ($MT$) or rotation ($CT$) combined with time reversal—must be preserved to strictly protect the zero net magnetization.

A straightforward approach to breaking the $PT$ symmetry is constructing an asymmetric FE/AFM bilayer [**Fig. 1(b)**]. While the interfacial electric polarization successfully lifts the momentum-space spin degeneracy, this simple Janus-like architecture suffers from a fatal physical flaw: the upward and downward FE states are not symmetrically equivalent. Consequently, reversing the polarization does not map the system onto a symmetric counterpart, rendering the resulting band structures and spin-splitting patterns analytically disjointed and highly unpredictable. Such thermodynamic and electronic asymmetry fundamentally precludes the deterministic and nonvolatile switching required for systematic device engineering.

To overcome this fundamental limitation, we formulate the symmetric FE/AFM/FE trilayer architecture, as illustrated in **Fig. 1(c)**. The out-of-plane polarization from the encapsulating FE layers robustly breaks the $PT$ symmetry of the central AFM layer, successfully generating the AM state. The transformative advantage of this trilayer geometry lies in its global symmetry: the opposite polarization states are now strictly equivalent. This equivalence arises because the two FE states are intrinsically linked by a global $PT$ operation. Specifically, the spatial inversion reverses the electric polarization and swaps the atomic coordinates while leaving the local spin moments invariant; subsequently, the time-reversal operation flips the magnetic moments, perfectly aligning the system into the reversed FE ground state. This rigorous $PT$ mapping mathematically guarantees that the momentum-resolved AM spin-splitting pattern is completely inverted upon polarization reversal. Furthermore, because the anomalous Hall conductivity ($\sigma_{xy}$) is parity-even and time-reversal-odd, this structural-magnetic transformation inherently enforces a pristine sign reversal of the macroscopic transport signal ($\sigma_{xy} \rightarrow -\sigma_{xy}$). By elegantly linking FE switching to an exact inversion of both the spin texture and transport signals, this symmetry-enforced equivalence establishes a universally predictable platform for altermagnetic control, completely bypassing the need for case-by-case material evaluations.

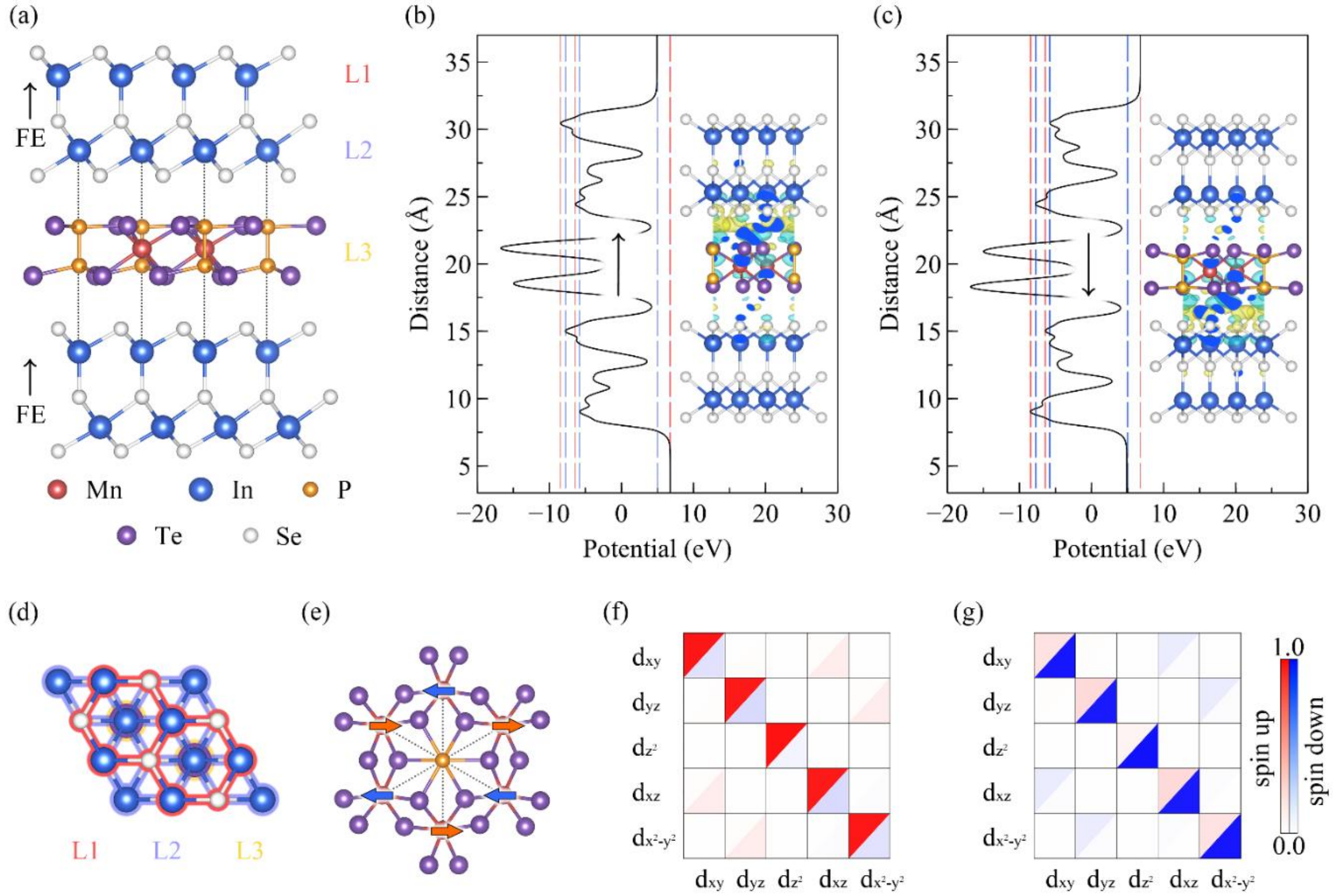

**Figure 2.** (a) Side and (d) top views of the optimized heterostructure in the FE-1 state. Black arrows denote the local FE polarization direction within the $In_2Se_3$ layers. Plane-averaged electrostatic potentials along the out-of-plane ($z$) direction for the (b) FE-1 and (c) FE-2 states. Insets display the corresponding differential charge densities, with yellow (orange) isosurfaces representing electron accumulation (depletion). The global polarization direction of the entire trilayer is indicated by macroscopic black arrows. (e) In-plane collinear AFM ground state of the central $MnPTe_3$ layer. Orbital-resolved occupation matrices for the $3d$ states of Mn atoms with magnetic moments aligned along the (f) $+x$ and (g) $-x$ directions. The spin quantization axis is fixed along the $x$-axis throughout the spin-up and spin-down definitions.

To physically realize this symmetry-enforced mechanism, we architect a concrete van der Waals heterostructure by interfacing the widely established AFM $MnPTe_3$ monolayer with two FE $In_2Se_3$monolayer [23–25]. The isolated $MnPTe_3$ monolayer crystallizes in a hexagonal lattice with the $P\overline{3}1m$ (No. 162) space group and a lattice constant of 6.98 Å [25]. Conversely, the $In_2Se_3$ monolayer adopts a rhombohedral structure with the $P3m1$ (No. 160) space group, with a lattice constant of 4.07 Å [24]. By assembling a $\sqrt{3} \times \sqrt{3}$ supercell of $In_2Se_3$ monolayer, we achieve a highly commensurate stacking with a negligible lattice mismatch of merely 0.99%, ensuring exceptional structural feasibility and thermodynamic stability.

In the assembled $In_2Se_3$/$MnPTe_3$/$In_2Se_3$ trilayer [**Fig. 2(a)** and **(d)**], the out-of-plane $C_3$ rotational axes of all the three constituent layers are perfectly aligned. Critically, the encapsulating $In_2Se_3$ layers explicitly break the $P$ symmetry. By maximally preserving the remaining symmetry operations, the entire heterostructure condenses into the $P31m$ (No. 157) space group. Upon full structural relaxation, the equilibrium lattice constant settles at 7.02 Å. Strikingly, the side profile [**Fig. 2(a)**] reveals a pronounced geometric distortion within the central $MnPTe_3$ layer, driven by the strong proximity effect of the encapsulating FE polarizations. Specifically, the P atomic pairs undergo a substantial rigid shift along the polarization axis. This stark geometric modulation unambiguously confirms the robust structural and electrostatic coupling between the FE layers and the active magnetic core.

Building upon this optimal trilayer architecture, we quantitatively evaluate its FE characteristic. The system exhibits two degenerate FE ground states, denoted FE-1 and FE-2, distinguished by their opposite out-of-plane polarization directions. Because the structural periodicity is intrinsically broken along the z axis, we determine the net electric dipole moment using the classical definition. The dual $In_2Se_3$ encapsulating layers yield a robust macroscopic polarization of 16.39 pC/m. This magnitude not only rivals that of isolated single-layer transition metal phosphorous chalcogenides (~8 pC/m) [26] but overwhelmingly surpasses the weak polarizations inherent to, such as bilayer $VSi_2P_4$(0.42 pC/m) [21], $MnPSe_3$ (0.14 pC/m) [10], and $PtBr_3$ (2.4 pC/m) [17]. The robust nature of this spontaneous polarization is further corroborated by the [**Fig. 2(b)** and **(c)**]. We observe a massive vacuum-level discontinuity of 1.85 eV across the heterostructure, manifesting as a giant built-in electrostatic field. Furthermore, the differential charge density analyses [inserts of **Fig. 2(b)** and **(c)**] reveal a striking asymmetry in the interfacial charge transfer: the charge redistribution is predominantly confined to the top $MnPTe_3$/$In_2Se_3$ interface in the FE-1 state, but strictly switches to the bottom interface in the FE-2 state. This stark interfacial contrast highlights the profound unpredictability of polarization-reversal effects in conventional, uncompensated FE/AFM bilayers, fundamentally reinforcing the necessity of our symmetric trilayer design.

Equally critical to the FE response is the underlying magnetic architecture. The central $MnPTe_3$ monolayer stabilizes an in-plane collinear AFM ground state [23], as depicted in **Fig. 2(e)**. The nominal valence electron configuration of Mn atoms is $3d^5 4s^2$; by donating two electrons to the chemical bonds, the remaining five electrons half-fill the five $d$ orbitals. This yield an expected magnetic moment of 5 $\mu_B$ per Mn atom, an exact localization that is explicitly confirmed by the orbital-resolved occupation matrices **[Figs. 2(f)** and **2(g)**]. Within the heterostructure, the encapsulating FE layers explicitly break the $M_z$ mirror symmetry and inversion symmetry. However, the vertical mirror plane perpendicular to the $xy$ plane is preserved. This specific spatial symmetry acts as the fundamental bridge between the two magnetic sublattices: it forces the macroscopic net magnetization to strictly vanish, yet safely permits the opposite spin channels to split vigorously in momentum space, thereby successfully crystallizing the altermagnetic phase.

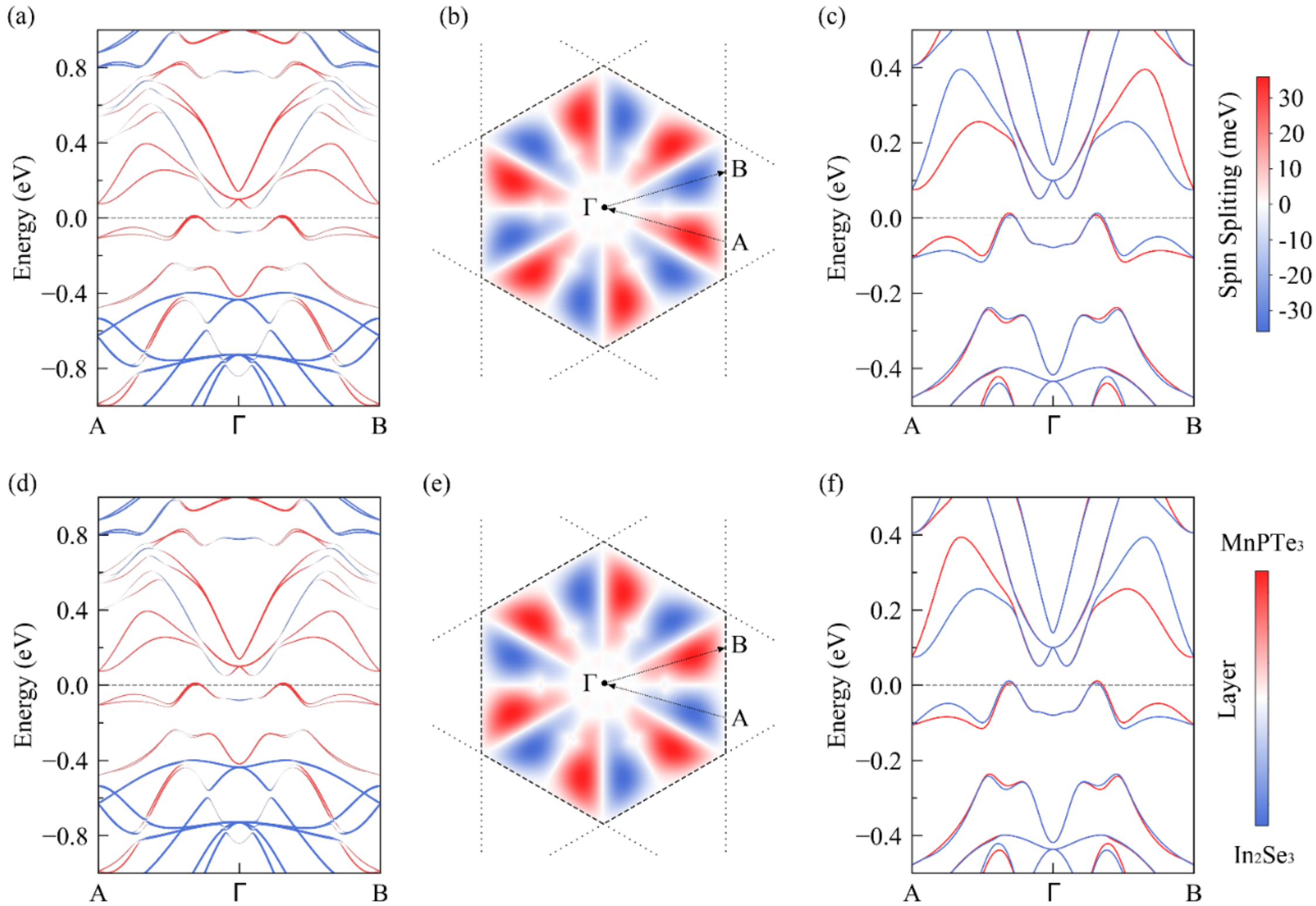

**Figure 3.** (a) Layer-resolved band structures of the FE-1 state. (b) Momentum-space colormap of the spin-splitting energy across the two-dimensional Brillouin zone for the FE-1 state. The dashed lines with arrows delineate the k-path used for the band dispersions. (c) Spin-resolved band structure of the FE-1 state. (d)–(f) The corresponding layer-projected bands, momentum-space spin-splitting colormap, and spin-resolved band structure for the reversed FE-2 state. The Fermi energy is shifted to zero.

To explicitly verify the emergence of the predicted altermagnetic state, we first examine the electronic band structures of the heterostructure. By projecting the electronic states onto the constituent layers [**Fig.**

**3(a)**], we reveal that the low-energy spectral weights near the Fermi energy are contributed mainly by the $MnPTe_3$ layer, with the encapsulating $In_2Se_3$ layers acting effectively as a wide-gap dielectric background. Notably, the profound interfacial charge transfer drives a semiconductor-to-metal transition within the $MnPTe_3$ monolayer [23], shifting the Fermi level deep into the active 3d electronic states of the $MnPTe_3$ layer. The spin-resolved band structure of the FE-1 state [**Fig. 3(c)**] exhibits a pronounced momentum-dependent spin splitting despite the strictly zero net macroscopic magnetization. This definitive lifting of the spin degeneracy explicitly confirms the crystallization of the altermagnetic phase.

To visualize the full anisotropic topography of this verified spin texture, we map the energy difference between the two spin channels across the entire Brillouin zone [**Fig. 3(b)**]. The spins degeneracy is protected along specific high-symmetry invariant momentum lines (e.g., Γ–M and Γ–K), manifesting as zero-splitting nodal axes. Moving away from these symmetry lines, the asymmetric crystal field vigorously breaks the local spin degeneracy, inducing a massive finite splitting that peak beyond 30 meV near the Brillouin zone boundaries. This highly anisotropic, momentum-locked spin texture fundamentally distinguishes the system from conventional ferromagnets, offering the tantalizing prospect of giant spin-polarized currents completely free from parasitic stray fields.

The definitive validation of our symmetric trilayer design lies in the deterministic control of this altermagnetic order via macroscopic ferroelectricity. By electrically reversing the out-of-plane polarization to access the FE-2 state, we physically invert the encapsulating electrostatic boundary conditions. Guided by our foundational symmetry principles, this spatial inversion swaps the two magnetic sublattices while preserving the vertical mirror symmetry. The consequence of this sublattice exchange is spectacularly confirmed by the spin-resolved band structures [**Fig. 3(f)**] and the corresponding Brillouin-zone colormap [**Fig. 3(e)**]: the altermagnetic spin-splitting pattern undergoes an exact, global inversion relative to the FE-1 state. Specifically, the local spin polarization at any generic momentum point $k$ is perfectly flipped, whereas the symmetry-enforced nodal degeneracies remain utterly invariant. This precise, deterministic mapping unambiguously corroborates our global $PT$-symmetry-enforced paradigm. It rigorously proves that reversing the FE polarization offers a nonvolatile, entirely predictable, and robust thermodynamic mechanism to toggle the altermagnetic spin texture, effectively establishing the $In_2Se_3/MnPTe_3/In_2Se_3$ architecture as a pristine material platform for electrically switchable altermagnetic spintronics.

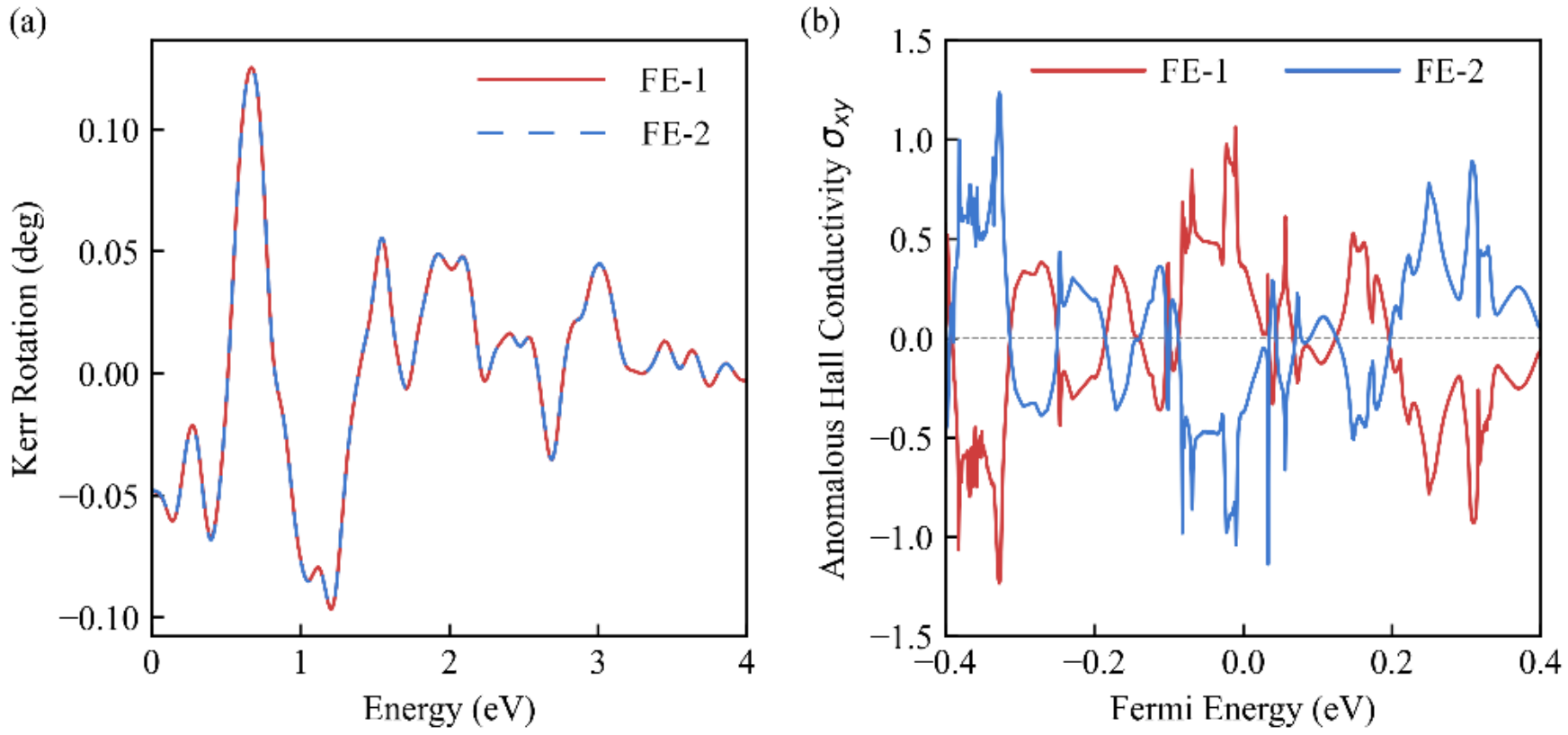

**Figure 4.** (a) Real parts of the Kerr signal as a function of photon energy for the FE-1 and FE-2 states. (b) Calculated anomalous Hall conductivity ($\sigma_{xy}$) versus the Fermi energy for both ferroelectric configurations. The Fermi energy is shifted to zero.

To experimentally probe this momentum-dependent spin splitting, we first evaluate the magneto-optical Kerr effect (MOKE), a highly sensitive optical diagnostic tool originating from the off-diagonal components of the optical conductivity tensor. The complex Kerr angle $\Phi_K = \theta_K + i\eta_K$ [27,28], which encapsulates both the Kerr rotation angle $\theta_K$ and ellipticity $\eta_K$, is evaluated via

$$\Phi_K = \frac{-\sigma_{yz}^{opt}}{\left(\sigma_{xx}^{\mathrm{opt}} - 1\right)\sqrt{\sigma_{xx}^{opt}}} \tag{1}$$

where the $\sigma_{xx}^{opt}$ and $\sigma_{yz}^{opt}$ are the diagonal and off-diagonal components of the optical conductivity tensor, respectively. As shown in **Fig. 4(a)**, the heterostructure exhibits a prominent Kerr peak of about 0.12° in the photon energy range of 0.5-1.0 eV, reflecting the significant altermagnetic splitting near the Fermi level. Crucially, our first-principles calculations show that the Kerr signal remains identical for both FE-1 and FE-2 states (see **Fig. 4a**). This invariance is fundamentally guaranteed by the crystal symmetry: while the two polarization states are physically distinct, they are mapped onto each other by a global $PT$ operation. Because the pertinent optical conductivity elements are invariant under this specific combined operation, MOKE inherently cannot distinguish between the two inverted altermagnetic orientations.

In contrast, the anomalous Hall effect (AHE) provides a definitive and macroscopic transport probe that unambiguously perfectly distinguishes the two FE states. We calculate the Berry curvature of the $In_2Se_3$/$MnPTe_3$/$In_2Se_3$ trilayer using the Wannier-interpolated tight-binding Hamiltonian and the Kubo formula [29]:

$$\Omega_{n,xy}(\boldsymbol{k}) = -2\,\mathrm{Im}\sum_{m\neq n}\frac{\langle u_{n\boldsymbol{k}}|\hat{v}_x|u_{m\boldsymbol{k}}\rangle\langle u_{m\boldsymbol{k}}|\hat{v}_y|u_{n\boldsymbol{k}}\rangle}{(\varepsilon_{m\boldsymbol{k}}-\varepsilon_{n\boldsymbol{k}})^2} \tag{2}$$

Here, $\hat{v}_{x/y}$ denotes the velocity operator, the indices $m$ and $n$ label the electronic bands, and $u_{n\boldsymbol{k}}$ and $\varepsilon_{n\boldsymbol{k}}$ are the periodic part of the Bloch wavefunction and the eigenvalue for band $n$ at wavevector $k$, respectively. The anomalous Hall conductivity （AHC） is then obtained by integrating the Berry curvature over the Brillouin zone [30]:

$$\sigma_{xy} = -\frac{e^2}{\hbar}\sum_n \int_{\mathrm{BZ}} \frac{d\boldsymbol{k}}{(2\pi)^3} f(\varepsilon_{n\boldsymbol{k}})\,\Omega_{n,xy}(\boldsymbol{k}) \tag{3}$$

The calculate AHC for the FE-1 and FE-2 states is plotted in **Fig. 4(b)**. As unequivocally dictated by macroscopic symmetry principles—since the AHC tensor $\sigma_{xy}$ is parity-even but time-reversal-odd—the $PT$ mapping between the two ferroelectric states intrinsically enforces a pristine and exact sign reversal ($\sigma_{xy} \rightarrow -\sigma_{xy}$). This completely invertible AHE signal spans the entire relevant energy window, thereby providing a highly robust, nonvolatile electrical fingerprint to experimentally address and distinguish the two equivalent altermagnetic spin-splitting states.

## Conclusion

In summary, we have established a universally applicable layer-engineering paradigm to achieve symmetry-enforced, nonvolatile electrical switching of two-dimensional altermagnetism. By embedding a conventional collinear AFM monolayer within a symmetric FE environment, we demonstrate that the spontaneous breaking of spatial inversion symmetry reliably generates robust momentum-dependent spin splitting. More fundamentally, because the two opposing FE states are strictly connected by global $PT$ symmetry, an electric-field-driven polarization reversal guarantees an exact, deterministic inversion of the entire altermagnetic spin texture. Rigorously validated in the $In_2Se_3$/$MnPTe_3$/$In_2Se_3$ van der Waals heterostructure, this exact thermodynamic mapping manifests macroscopically as a perfectly invertible anomalous Hall conductivity, offering an unambiguous transport readout. By decisively circumventing the stringent symmetry constraints of single-phase multiferroics and overcoming the inherently weak polarization of sliding mechanisms, our symmetric trilayer architecture provides a highly efficient, robust, and versatile physical framework for realizing the next generation of electrically programmable altermagnetic spintronics.

## Method

Our first-principles calculations are carried out using the Vienna ab initio simulation package (VASP) [31,32] within the framework of density functional theory (DFT) [33]. The Perdew-Burke-Ernzerhof (PBE) [34] functional within the generalized gradient approximation (GGA) is employed to treat the exchange-correlation potential. To eliminate spurious interactions between periodic images along the out-of-plane direction, a vacuum layer of 15 Å is inserted. A plane-wave energy cutoff of 500 eV and a 9×9×1 Gamma-centered k-point mesh for the primitive unit cell is adopted. All atomic positions and lattice constants are fully relaxed until the residual forces are less than 0.01 eV/Å, and the electronic self-consistent convergence criterion is set to $10^{-6}$ eV. Following the previous work, the on-site Coulomb interaction of Mn-3d electrons is taken into account using the GGA+U method [35] with an effective Hubbard parameter U=5 eV [36]. To obtain accurate band structures and Berry curvature, the maximally localized Wannier functions are constructed using the Wannier90 code [37], from which the real-space tight-binding Hamiltonian $H_R$ is extracted.

## Acknowledgments

This work is supported by the National Natural Science Foundation of China (No. 12274261), Taishan Scholar Program of Shandong Province, the Shandong Provincial Natural Science Foundation of China (ZR2024QA016), and Shandong Province Fundamental Research Zone (TQ012025003).

## Notes

During the completion of our work, we became aware of a related study by Chen et al. [38], which addresses a similar phenomenon but in a different material system.